\RequirePackage{fixltx2e}%required to be placed here in order to properly include figures

\documentclass[aip,jap, floatfix,reprint,groupedaddress,showpacs,showkeys,amsmath,amssymb,]{revtex4-1}
\usepackage{amsfonts}%
\usepackage[ngerman,english]{babel}
\usepackage[utf8]{inputenc}
\usepackage{graphicx}
\usepackage{dcolumn}% Align table columns on decimal point
\usepackage{bm}% bold math
\usepackage{color}

\begin{document}

%\preprint{}

\title{Electronic excitations and structure of Li\textsubscript{2}IrO\textsubscript{3} thin films grown on ZrO\textsubscript{2}:Y~(001) substrates}


\author{Marcus Jenderka}
	\email[Corresponding Author: ]{marcus.jenderka@physik.uni-leipzig.de}
\author{Rüdiger Schmidt-Grund}
\author{Marius Grundmann}
\author{Michael Lorenz}
	
%\thanks{}
\affiliation{%
Institut für Experimentelle Physik II, Universität Leipzig\\
Linnéstraße 5, D-04103 Leipzig (Germany)
}%


\date{\today}

\begin{abstract}
Thin films are a prerequisite for application of the emergent exotic ground states in iridates that result from the interplay of strong spin-orbit coupling and electronic correlations. We report on pulsed laser deposition of Li$_{2}$IrO$_{3}$ films on ZrO$_{2}$:Y(001) single crystalline substrates. X-ray diffraction confirms preferential (001) and (10-1) out-of-plane crystalline orientations with well defined in-plane orientation. Resistivity between 35 and 300~K is dominated by a three-dimensional variable range hopping mechanism. The dielectric function is determined by means of spectroscopic ellipsometry  and, complemented by  Fourier transform infrared transmission spectroscopy, reveals a small optical gap of $\approx$~300~meV, a splitting of the $5d$-$t_{2g}$ manifold, and several in-gap excitations attributed to phonons and possibly magnons.

\end{abstract}


\maketitle

%\section{}
%\subsection{}
%\subsubsection{}

%-------------Introduction

\section{Introduction}
The layered perovskite oxides A$_{2}$IrO$_{3}$ (A = Na, Li) have in recent years been studied in terms of a physical realization of the Kitaev and Heisenberg-Kitaev model and its extensions, harboring spin liquid and topologically ordered phases. \cite{Kitaev2006,Jackeli2009,Chaloupka2010,Reuther2011,Singh2012,Reuther2014a} They have also drawn interest as possible topological insulators. \cite{Shitade2009,Pesin2010,Kim2012,Kim2013} Ultimately, the physical realization of such states of matter is desired with respect to quantum computation proposals. \cite{Kitaev2003,Collins2006,Nayak2008} Subsequently, experiments showed that both materials order magnetically at $\approx$ 15 K: Na$_{2}$IrO$_{3}$ is deep within an antiferromagnetically zig-zag ordered phase, \cite{Singh2010,Ye2012,Liu2011,Choi2012} whereas Li$_{2}$IrO$_{3}$ shows incommensurate spiral order but is placed close to the desired spin liquid phase. \cite{Kobayashi2003,Singh2012,Reuther2014a} There is yet no direct experimental evidence of a topological insulator phase in either of the two materials.

Some of us have previously reported on the first successful growth of heteroepitaxial Na$_{2}$IrO$_{3}$ thin films, where we observed three-dimensional variable range hopping conductivity and the weak antilocalization effect in magnetoresistance. \cite{Jenderka2013} To date, available single crystals are of very small size, such that experimental data on Li$_{2}$IrO$_{3}$ is often restricted to powder-averaged data. \cite{Kobayashi1997,Kobayashi2003,Singh2012,Gretarsson2013} Certain types of experiments, such as terahertz pump-probe spectroscopy or neutron diffraction, benefit from large-area single-crystalline thin film samples. Neutron diffraction experiments, for instance, are challenging because of a high absorption cross-section of iridium and the small crystal sizes. Thin films however alleviate these problems by distributing the volume over a large area. If its growth was feasible, thin films would in general also allow for the study of strain-induced effects. In fact, both a strain-induced spin liquid and topological insulator phase are proposed to exist. \cite{Reuther2011,Singh2012,Kim2012,Kim2013} Hence, thin films of this class of material lay a foundation for future experiments including e.g. studies on the (Na$_{1-x}$Li$_x$)$_2$IrO$_{3}$ compounds, which are not available in single crystalline form so far.

Like its sister compound Na$_{2}$IrO$_{3}$, Li$_{2}$IrO$_{3}$ is an antiferromagnetic insulator with Ne\'{e}l temperature $T_{\mathrm{N}}$~$\approx$~15~K below which it orders in an incommensurate spiral fashion. \cite{Kobayashi2003,Singh2012, Reuther2014a} X-ray powder diffraction measurements suggest a monoclinic C2/c unit cell. \cite{Kobayashi2003,Singh2012} Temperature dependent resistivity shows insulating behavior between 100 and 300 K. \cite{Singh2012} A delicate interplay between trigonal distortions of the IrO$_6$ octaeder and spin-orbit coupling cause the spin-orbit assisted Mott insulating state. Until recently, the magnitude of the trigonal distortions was unclear. Consequently, the adequate description of the underlying electronic structure was under debate. \cite{Mazin2012a,Mazin2013,Bhattacharjee2012} However, recent resonant inelastic x-ray scattering experiments \cite{Gretarsson2013} validate the applicability of the so-called $j_{\mathrm{eff}}$ physics in Li$_{2}$IrO$_{3}$, which in the past was applied to Sr$_2$IrO$_4$ and related materials \cite{Kim2008,Moon2008,Kuriyama2010,Lee2013a}: The cubic crystal field caused by the edge-sharing IrO$_6$ octaeder splits the Ir 5$d$ orbitals into a $e_{\mathrm{g}}$ doublet and a $t_{\mathrm{2g}}$ triplet. The splitting between $e_{\mathrm{g}}$ and $t_{\mathrm{2g}}$ is about 3 eV. \cite{Moon2006} Structural distortions of the monoclinic unit cell and the oxygen tetraeder, respectively create a trigonal crystal field. If this field is sufficiently small compared to spin-orbit coupling, the $t_{\mathrm{2g}}$ triplet further splits into a fourfold degenerate $j_{\mathrm{eff}}$ = 3/2 and a twofold degenerate $j_{\mathrm{eff}}$ = 1/2 band.

In this paper, we report on the pulsed laser-deposition of Li$_2$IrO$_3$ thin films and the study of their structure. Temperature-dependent resistivity is measured between 300 and 25 K. Employing spectroscopic ellipsometry, we determine the dielectric function  in order to study and interpret low energy electronic excitations on the basis of $j_{\mathrm{eff}}$ physics. These results are complemented with optical transmission data measured with a Fourier transform infrared spectrometer.

%-------------------------------Experimental details-----------------------------------------
%
\section{Experimental details}

Li$_2$IrO$_3$ thin films were grown by pulsed laser depositon (PLD) on 10 $\times$ 10  mm$^\mathrm{2}$ ZrO$_2$:Y(001) single crystals (YSZ). PLD was done with a 248 nm KrF excimer laser at a laser fluence of 2 Jcm$^{\mathrm{-2}}$.  The phase-pure polycrystalline target was prepared by a solid state synthesis of Li$_{2}$CO$_{3}$ and IrO$_{2}$ powders in a stochiometric ratio of 1.1:1. The mixture was homogenized, pressed and calcined in air for 24 h at 750$^\circ$C. Afterwards it was again ground, pressed and sintered for 72 h in 900$^\circ$C. Unfortunately, the target was rather soft. To improve it, attempts were made to employ high pressure sintering (HPS). However, the reducing conditions due to the graphite heater used in HPS, had a detrimental effect on the target composition.

The deposition procedure involved a nucleation layer grown with 300 laser pulses at 1 Hz, followed by another 30,000 pulses at 15 Hz. After deposition, the samples were annealed in situ at an oxygen partial pressure $p_{\mathrm{O2}}$~=~800~mbar. Film thickness was estimated at $\approx$~400~nm by spectroscopic ellipsometry. The growth process was optimized with respect to high film crystallinity, i.e. high x-ray diffraction peak intensity. We have grown films  at growth temperatures and oxygen partial pressures within the ranges from 550$^\circ$C to 700$^\circ$C, and from 0.1~mbar to 3.0~$\times$~10$^{\mathrm{-4}}$ mbar, respectively. We note, that for larger deposition temperatures Li$_2$IrO$_3$ films decomposed as indicated by additional x-ray reflexes belonging to other phases, most likely Ir$_x$O$_y$ species. For these films, an energy dispersive x-ray (EDX) analysis gave a reduced Ir:O ratio of 1:2.13. Optimized films were grown at 600$^\circ$C  and 3.0 $\times$ 10$^{\mathrm{-4}}$ mbar.

Investigations of the epitaxial relationship were performed with both a Panalytical X’Pert PRO Materials Research Diffractometer with parabolic mirror and PIXcel$^{3D}$ detector and a Philips X'Pert x-ray diffractometer equipped with a Bragg-Brentano powder goniometer using divergent/focusing slit optics and Cu K$_{\mathrm{\alpha}}$ radiation. Surface morphology was investigated via a Park System XE-150 atomic force microscope in dynamic non-contact mode and a CamScan CS44 scanning electron microscope. Temperature dependent dc electrical resistivity was measured in van-der-Pauw geometry with dc magnetron-sputtered ohmic gold contacts. The dielectric function (DF) was determined via standard variable angle spectroscopic ellipsometry in the spectral range from 0.03 to 3.34 eV (IRSE). Infrared optical transmission in the spectral range from 0.14 to 1.30 eV was measured using a BRUKER IFS 66v/S Fourier transform infrared spectrometer in transmission mode (T-FTIR).

%---------------------------------Results---------------------------------------------------------
\section{Results \& discussion}
\subsection{Crystal structure}
%
%-------------Figure: Structure---------------------------------
\begin{figure}
\includegraphics[width=\columnwidth]{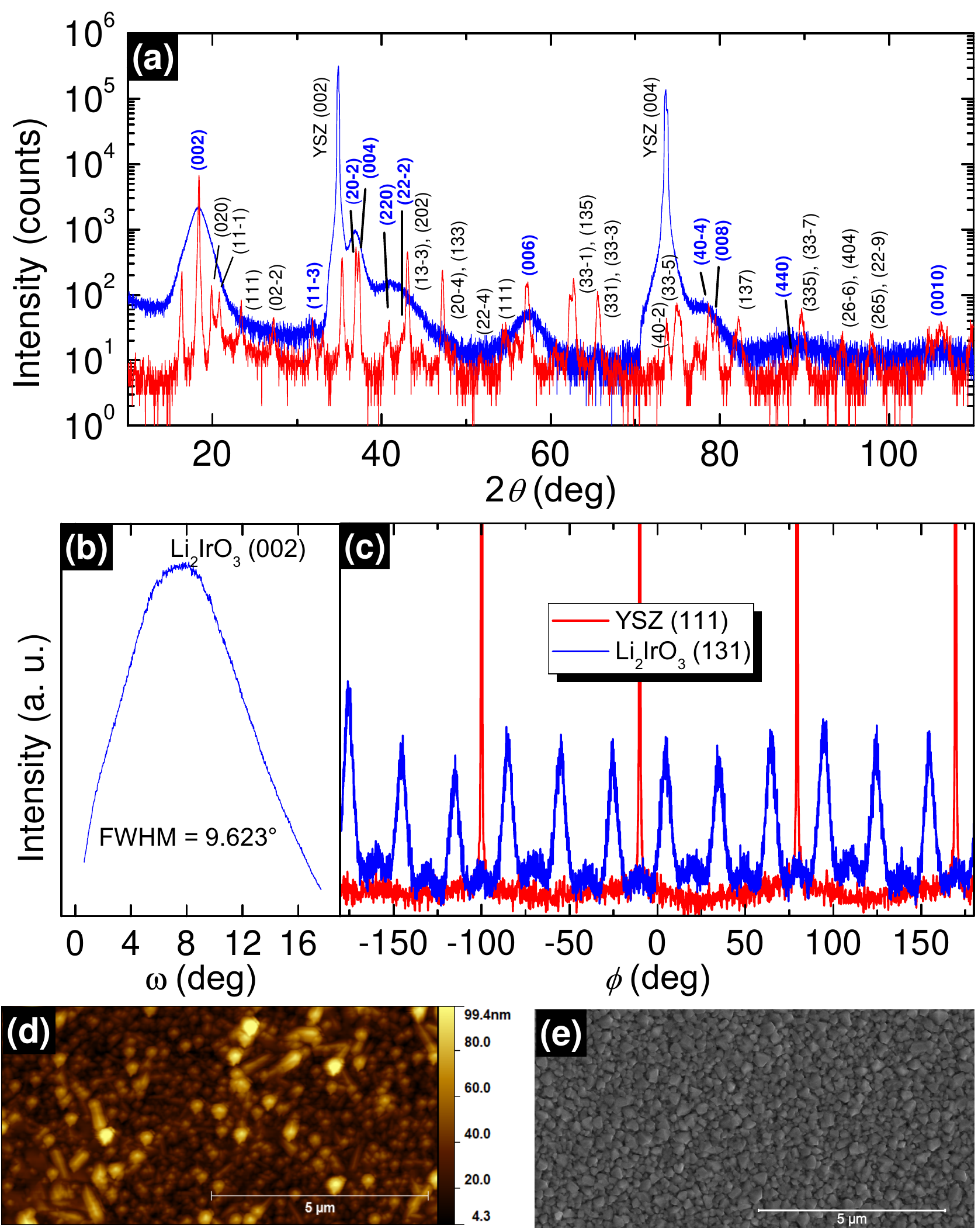}%
\caption{\label{fig:structure}(Color online) (a) Typical x-ray diffraction (XRD) \mbox{2$\Theta$-$\omega$} scan of a PLD-grown Li$_{2}$IrO$_{3}$ thin film on ZrO\textsubscript{2}:Y~(001) (YSZ) in blue, underlayed with the pattern of the polycrystalline PLD target in red. Significant film reflexes are labeled in blue. (b) Rocking curve about the (002) plane of Li$_2$IrO$_3$. (c) XRD $\phi$-scans of asymmetric Li$_{2}$IrO$_{3}$ (131) and YSZ (111) reflections of the Li$_{2}$IrO$_{3}$(001) phase. (d,e) The surface morphology of the film is illustrated by non-contact AFM topographic and scanning electron microscopy images.}
\end{figure}
A typical x-ray diffraction (XRD) 2$\Theta$-$\omega$ pattern of a Li\textsubscript{2}IrO\textsubscript{3} film is shown in Fig. \ref{fig:structure}(a). For confirmation of the preferential film orientations, the XRD pattern of the polycrystalline PLD target is underlayed. The patterns are indexed on the basis of a monoclinic $C2/c$ unit cell \cite{Kobayashi2003,Singh2012}. The pattern shows pronounced symmetric peaks related to the (001) and (10-1) planes of the Li\textsubscript{2}IrO\textsubscript{3} phase, see blue labels in Fig. \ref{fig:structure}(a). The intensity ratio $I_{(002)}$/$I_{(20-2)}$ is about 2.3:1 and thus much smaller than in the polycrystalline target ($\approx$~14:1). This strongly suggests preferential (001) and (10-1) out-of-plane crystalline orientations. However, peak assignment can be ambiguous, since in the monoclinic unit cell of Li$_2$IrO$_3$ many reflexes such as (20-2) \& (004) are narrowly spaced. We fit the peak at $2\Theta$~=~36.88$^\circ$ with two Gaussians to obtain $d_{\mathrm{20-2}}$ and $d_{\mathrm{004}}$. From the $d_{\mathrm{00l}}$ values of the (002), (004) and (006) reflexes, the $c$ lattice parameter is determined as 9.71(1)~$\mathrm{\mathring{A}}$. We also performed $2\Theta$-$\omega$ scans of the asymmetric (202) and (131) reflexes of the (001)-oriented phase and calculated lattice parameters $a$~=~5.20(3)~$\mathrm{\mathring{A}}$ and $b$~=~8.99(14)~$\mathrm{\mathring{A}}$. For the calculations we assumed $\beta$~=~99.992$^\circ$. \cite{Kobayashi2003} The lattice parameters are in good agreement with reported values \cite{Kobayashi2003,Singh2012}, $a$, $b$ and $c$ deviating by +0.6\%, +0.7\% and -0.8\% respectively. This indicates some amount of internal rather than epitaxial strain considering the significant lattice mismatch to YSZ(001).

A rocking curve of the (002) reflex at $2\Theta$~=~18.29$^\circ$ gives a full width at half maximum of about 10$^\circ$ (cf. Fig \ref{fig:structure}(b)).

Investigation of the in-plane epitaxial relationship is done by x-ray diffraction $\phi$-scans of the asymmetric Li\textsubscript{2}IrO\textsubscript{3} (131) reflex of the (001)-oriented phase and of YSZ (111), shown in Fig. \ref{fig:structure}(c). For YSZ(001), we observe the C$_{\mathrm{4}}$-symmetry of its (111) planes as expected. The $\phi$-scan of the Li\textsubscript{2}IrO\textsubscript{3} (131) reflection shows 12~+~12 reflections spaced by 15$^{\circ}$ alternating between "high" and "low" intensity peaks. Their intensity ratio is approximately 7:1. In Li$_2$IrO$_3$, the asymmetric (131) plane of both the (001)- and (10-1)-oriented phases share very similar angles. We therefore believe, that the $\phi$-scan shows the rotational domains of both phases. Thus, it indicates that the C$_{\mathrm{1}}$-symmetric (monoclinic) (001)- and (10-1) oriented Li\textsubscript{2}IrO\textsubscript{3} epilayers align in-plane within 12 rotational domains each. From the mismatch of rotational symmetry of substrate and epilayer,  \cite{Grundmann2010} we expect a minimum of four rotational domains. The origin of the increased number of rotational domains is at present unknown. One explanation might however be possible: Initially, Li$_2$IrO$_3$ was reported as pseudo-hexagonal. \cite{Kobayashi1997} If it occurred, the existence of epitaxially induced C$_{\mathrm{6}}$-symmetry would lead, according to Ref. \onlinecite{Grundmann2010}, to two rotational domains and thus 12 reflections in total per phase. Based on our x-ray data, it is not possible to determine the actual crystal structure of the epilayer, as the (001) planes of both monoclinic and pseudohexagonal symmetries share similar $2\Theta$ angles. Epitaxially induced change of lattice symmetry was e.g. observed for the growth of U$_3$O$_8$ on c-plane Al$_2$O$_3$.~\cite{Burrell2007}

\subsection{Chemical analysis and surface morphology}
A qualitative elemental investigation of a film using secondary neutral mass spectrometry (SNMS) showed only Li, Ir, and O. An accumulation of volatile Li at the film-substrate interface was observed by SNMS depth profiling. However, it is currently not clear if the Ar$^{+}$-sputtering in SNMS contributes to this Li accumulation at the interface. Energy dispersive x-ray analysis gave an oxygen deficient Ir:O ratio of 1:2.84 under the optimized PLD conditions, as discussed above. This ratio points to possible oxygen vacancies, that might in turn explain the internal strain observed in XRD.

Figures \ref{fig:structure}(d,e) show topographic images of the Li\textsubscript{2}IrO\textsubscript{3} film surface obtained with non-contact atomic force microscopy (AFM) and scanning electron microscopy (SEM), respectively. The images reveal a granular surface with an RMS roughness of 15.7 nm and a peak-to-valley height of 139.0 nm. The surface morphology can be explained by the presence of two preferential orientations and furthermore by the mechanically soft target promoting grain agglomeration. We have in fact tried growing films on lattice-matched YAlO$_3$(011) single crystals but observed no improvement in either crystalline structure or surface morphology.

\subsection{Electrical resistivity}
\begin{figure}
\includegraphics[width=\columnwidth]{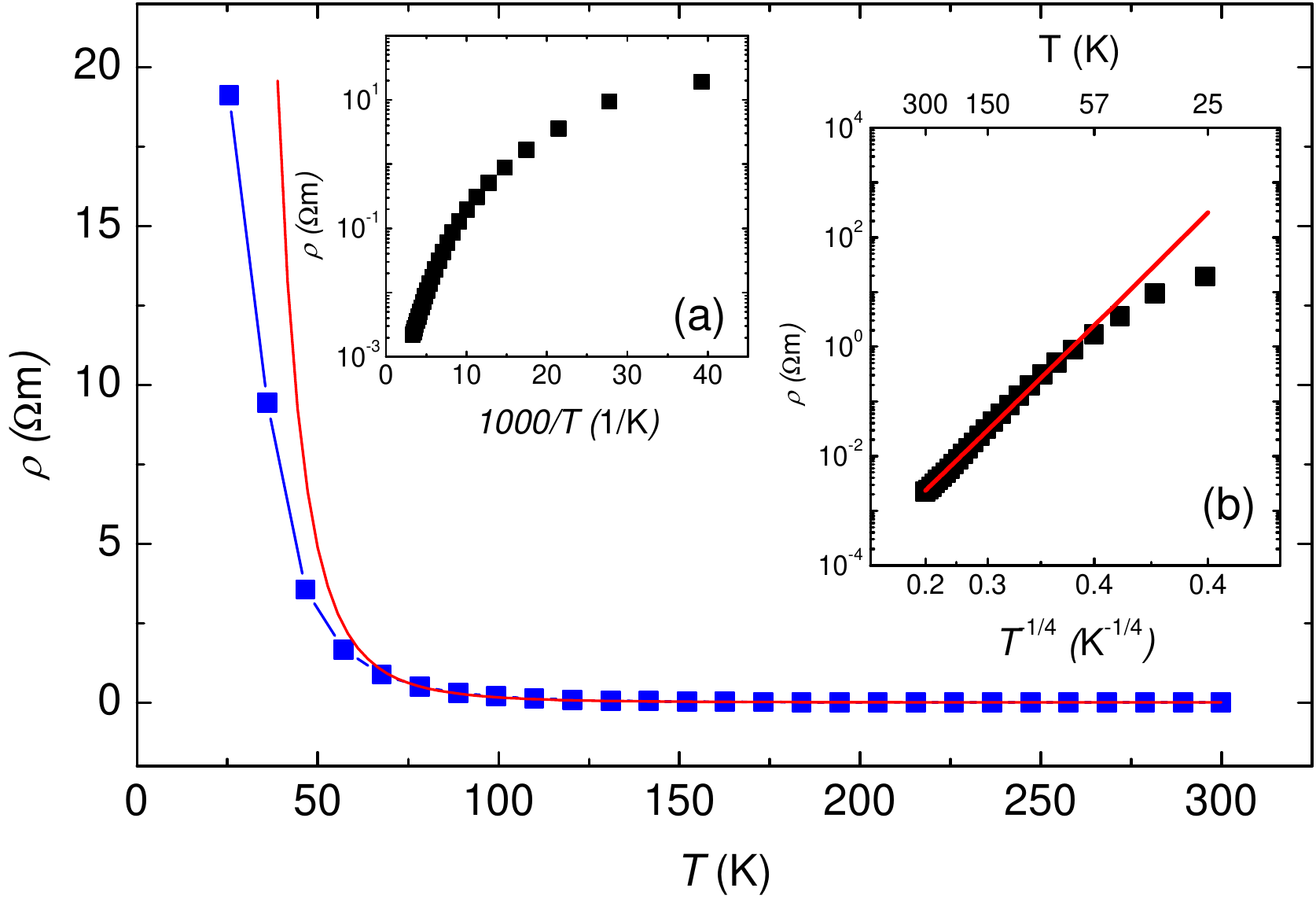}%
\caption{\label{fig:res}(Color online) Temperature-dependent resistivity $\rho$ versus temperature $T$. The red line is a fit according to eq.~(\ref{eq:MottVRH}). Inset (a) shows log$\rho$ versus $1000/T$. In inset (b) the data are plotted in log$\rho$ versus $T^{-1/4}$. Red straight line fit with slope $T_{\mathrm{0}}$ illustrates dominant three-dimensional Mott variable range hopping conductivity mechanism.}
\end{figure}

As illustrated in Fig. \ref{fig:res}, Li$_{2}$IrO$_{3}$ exhibits semiconducting resistivity behavior between 300 and 25 K. Resistivity does not follow a simple activated law $\rho \propto \exp(-\Delta/T)$, see Fig. \ref{fig:res}, inset (a), that can also be associated with nearest-neighbor hopping. Instead, resitivity follows a $\rho \propto \exp[(\Delta/T^{1/4})]$ dependence down to at least 60 K. We associate this behavior to three-dimensional Mott variable range hopping (VRH), as similarly observed in Na$_{2}$IrO$_{3}$ thin films. \cite{Jenderka2013} Variable range hopping conduction results from localized states within a narrow energy band near the Fermi energy. At lower temperatures, resistivity diverges from this VRH dependence. As indicated by a straight line fit in Fig. \ref{fig:res}, inset (b), we fit the resistivity data from 300 to 57 K with the three-dimensional Mott VRH model \cite{Mott1969}
\begin{equation}
\rho = \rho_{\mathrm{0}}\exp[(T_{\mathrm{0}}/T)^{1/4}],
\label{eq:MottVRH}
\end{equation}
where $\rho_{\mathrm{0}}$ is a temperature-dependent resisitivty coefficient and $T_{\mathrm{0}}$ is the localization temperature. From the analysis of variable range hopping within percolation theory, the resisitivty coefficient is given by \cite{Shklovskii1984,Arginskaya1994a}
\begin{equation}
\rho_{\mathrm{0}} =  \rho_{\mathrm{0}}' (T/T_{\mathrm{0}})^{s},
\end{equation}
where $s\approx$ 1/4. Fitting eq. (\ref{eq:MottVRH}) to the resistivity data, we extract $\rho_{\mathrm{0}}'$~=~2.0(4)~$\times$~10$^{-8}$~$\mathrm{\Omega m}$  and a localization temperature of $T_{\mathrm{0}}$~=~1.26(6)~$\times$~10$^{7}$~K. An estimate for the localization length $a$ can be obtained from $T_{\mathrm{0}}$ according to \cite{Shklovskii1984}
\begin{equation}
T_{\mathrm{0}}=21.2/k_{\mathrm{B}}a^{3}N(E_{\mathrm{F}}),
\end{equation}
provided that $N(E_{\mathrm{F}})$, the density of states at the Fermi level, is known. Based on available heat capacity data of Li$_2$IrO$_3$ and Na$_2$IrO$_3$ single crystals \cite{Singh2012}, we assume $N(E_{\mathrm{F}})$~$\sim$10$^{28}$~eV$^{-1}$m$^{-3}$ (for details, see Ref. \onlinecite{Jenderka2013}). The calculated localization length $a$ is about 1.25(2) \AA. Its magnitude is comparable with the Ir-Ir and Ir-O bond distances (approx. 3 and 2 \AA, resp.) in the structurally very similar sister compound Na\textsubscript{2}IrO\textsubscript{3}, \cite{Choi2012,Ye2012} supporting the applicability of Mott VRH. A criterion for the validity of Mott VRH is that the average hopping distance $R_{\mathrm{M}}$ be larger than the localization length $a$, \cite{Mott1968} i.e.
\begin{equation}
\frac{R_{\mathrm{M}}}{a} = \frac{3}{8}\left(\frac{T_{\mathrm{0}}}{T}\right)^{1/4}>1.
\label{eq:Rmean}
\end{equation}
The ratio $R_{\mathrm{M}}$/$a$ is equal to 5.37(7) at 300 K satisfying above criterion. We also verify that the maximal hopping distance $R_{\mathrm{max}}$ is much larger than the film thickness to exclude Mott VRH with hopping exponent 1/3. This kind of hopping is expected for thin films with thickness $t$ in the order of the maximal hopping distance \cite{Shante1973}
\begin{equation}
R_{\mathrm{max}}= \frac{a}{2}\left(\frac{T_{\mathrm{0}}}{T}\right)^{1/4}.
\end{equation}
Using the estimated localization length $a$ we obtain a maximal hopping distance of $R_{\mathrm{max}}$ = 1.36(4) nm at 57 K which is much smaller than the film thickness $t$ $\approx$~400~nm.

%Ellipsometrie
\subsection{Spectroscopic ellipsometry}
\begin{figure}
\includegraphics[width=\columnwidth]{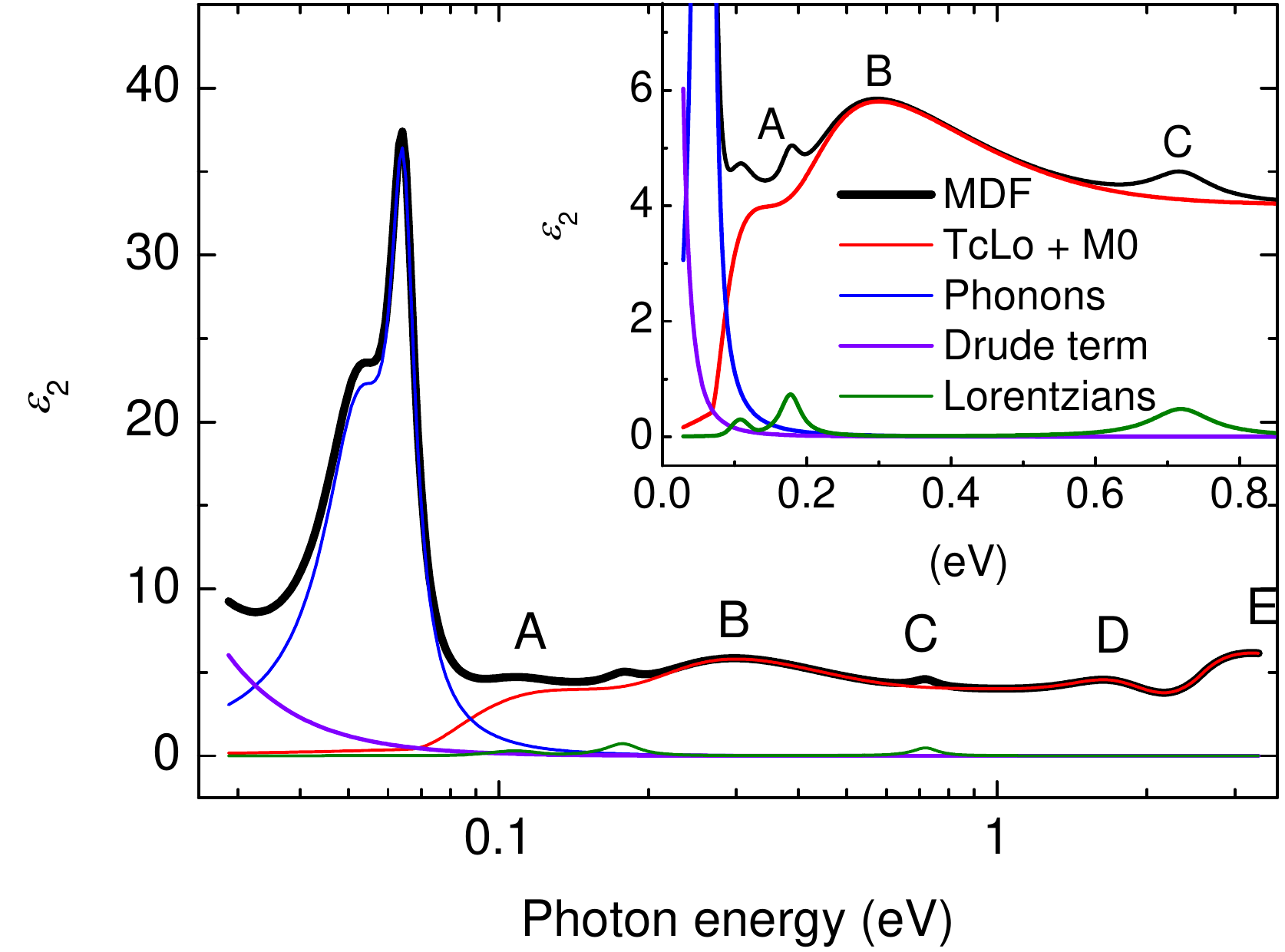}%
\caption{\label{fig:IRSE}(Color online) Model dielectric function (MDF) and its individual components as obtained by spectroscopic ellipsometry of a Li$_2$IrO$_3$ thin film on YSZ(001). Labels $A$ to $E$ denote electronic excitations, see text and Tab. \ref{tab:excitations}. The inset shows a zoom-in on the electronic excitations below 0.8 eV.}
\end{figure}

Spectra of the DF were determined from the ellipsometry data by using a model containing a layer for the substrate, a layer for the Li$_2$IrO$_3$ film and a surface roughness layer. Each layer is described by its thickness and optical constants. Between the substrate and the Li$_2$IrO$_3$ thin film, a thin interface layer of about 1 nm was introduced by mixing 50:50 the substrate and film DF in a ratio of 1:1 by means of a Bruggeman effective-medium approximation (EMA). \cite{Jellison1994} Also the surface roughness layer and the columnar structure was accounted for by a 100 nm EMA-layer by mixing the film DF with void. The void fraction was gradually increased from $\approx 20$ to $\approx$ 80\% from bottom to top (please cf. note \footnote{Please note that such a high surface roughness cannot be exactly described by this approach and the DF of the surface layer with their chosen gradient strongly correlates with the DF of the thin film. Therefore, for absolute values of the thin film DF a uncertainty of $\Delta \varepsilon_2 \approx +20\% / -5\%$ in the worst case has to be taken into account. But nevertheless, their general lineshape and the energy values of the electronic excitations and phonon modes are influenced far less and are therefore reliable. For the same reason we do not report $\varepsilon_1$ because here the error amounts up to 50\%.}). For modelling the thin film's DF we used a parametric model dielectric function (MDF) approach consisting of (cf. Fig. \ref{fig:IRSE}): a Drude free charge carrier absorption and two Lorentzians describing the phonon contribution at lowest energies; a near band gap Tauc-Lorentz (TcLo) absorption function \cite{Jellison1996} and a series of M0-critical point functions with parabolic onset \cite{Adachi1987,Yoshikawa1997} describing direct band-band transitions; Gaussian oscillators were used to model electronic band-band transitions spread within the Brillouin zone at higher energies. Additional discrete transitions where described by Lorentzians. Regression analysis was then applied to best match the dielectric function model to the experimental data. From the MDF, the absorption coefficient $\alpha$ was calculated (cf. Fig. \ref{fig:FTIR}).

The final MDF together with its individual components is displayed in Fig. \ref{fig:IRSE}. In the following, we argue that these individual components represent electronic excitations that can be ascribed to transitions from and within the $t_{\mathrm{2g}}$ manifold in line with the picture of $j_{\mathrm{eff}}$ physics, as explained earlier in the text. We also show, that their transitions energies agree well with other experiments, such as angle-resolved photoemission spectroscopy (ARPES), resonant inelastic x-ray scattering (RIXS) and optical spectroscopy, performed on Li$_{2}$IrO$_{3}$, Na$_{2}$IrO$_{3}$ and other iridates. The observed excitations are summarized in Table \ref{tab:excitations}.

In the high-energy spectral range above 1.2 eV, we find two contributions to the MDF at 1.63 and 3.25 eV that are attributed to $d$-$d$ band transitions from occupied $t_{\mathrm{2g}}$ to empty $e_{\mathrm{g}}$ bands ($D$, $E$). Very similar transitions have been observed in Na$_{2}$IrO$_{3}$ and Li$_{2}$IrO$_{3}$ single crystals \cite{Comin2012,Gretarsson2013} and Na$_{2}$IrO$_{3}$ thin films. \cite{Jenderka2013} Below 1.2 eV, we find several transitions. The transitions at 55 and 65 meV are attributed to phonons; at 0.11 and 0.18 eV we find discrete transitions ($A$). For example, in Sr$_2$IrO$_4$ a magnon was found at 0.2 eV within the Mott gap. \cite{Kim2012b} Further band-band excitations, possibly related to the particle-hole continuum boundary, \cite{Gretarsson2013} are found at 0.15 and more pronounced at 0.30 eV ($B$). They imply a very narrow Mott gap of less than 0.3 eV. Narrow Mott gaps from 300 meV to 340 meV were also previously found in Na$_2$IrO$_3$ and Li$_2$IrO$_3$. \cite{Comin2012,Gretarsson2013,Jenderka2013} Another discrete transition can be recognized at 0.72 eV ($C$). It is assigned to $t_{\mathrm{2g}}$ intraband transitions from $j_{\mathrm{eff}}$ = 3/2 to $j_{\mathrm{eff}}$ = 1/2 states. Excitations similar to $C$ were found in other iridates, as well. \cite{Moon2008,Kim2012b,Gretarsson2013}
%
%-------------Figure: ellipsometry----------------------
\begingroup
\begin{table*}
\caption{\label{tab:excitations}%
Overview of electronic excitations energies in Li$_2$IrO$_3$ thin films determined by IRSE and T-FTIR. The uncertainty is in the order of the last digit.}
\begin{ruledtabular}
\begin{tabular}{lccccccc}
Peak  & \textit{A} & \textit{B} & \textit{B'} & \textit{C} & \textit{C'} & \textit{D} & \textit{E} \\
\colrule
Energy (eV) & 0.11, 0.18 & 0.3   & 0.43  & 0.72  & 0.89  & 1.63  & 3.25 \\
\colrule
type of & magnon & $j_{\mathrm{eff}}$ = 3/2  $\rightarrow$ & $j_{\mathrm{eff}}$ = 3/2  $\rightarrow$ & $j_{\mathrm{eff}}$ = 3/2  $\rightarrow$ & $j_{\mathrm{eff}}$ = 3/2  $\rightarrow$  & $t_{\mathrm{2g}}$ $\rightarrow$ $e_{\mathrm{g}}$ & $t_{\mathrm{2g}}$ $\rightarrow$ $e_{\mathrm{g}}$ \\
excitation & (discrete) & $j_{\mathrm{eff}}$ = 1/2 & $j_{\mathrm{eff}}$ = 1/2 & $j_{\mathrm{eff}}$ = 1/2 & $j_{\mathrm{eff}}$ = 1/2  & (band-band) & (band-band) \\
 & & (discrete) & (band-band) & (discrete) & (discrete) &  &  \\
\end{tabular}
\end{ruledtabular}
\end{table*}
\endgroup

\subsection{Infrared transmission spectroscopy}
\begin{figure}
\includegraphics[width=\columnwidth]{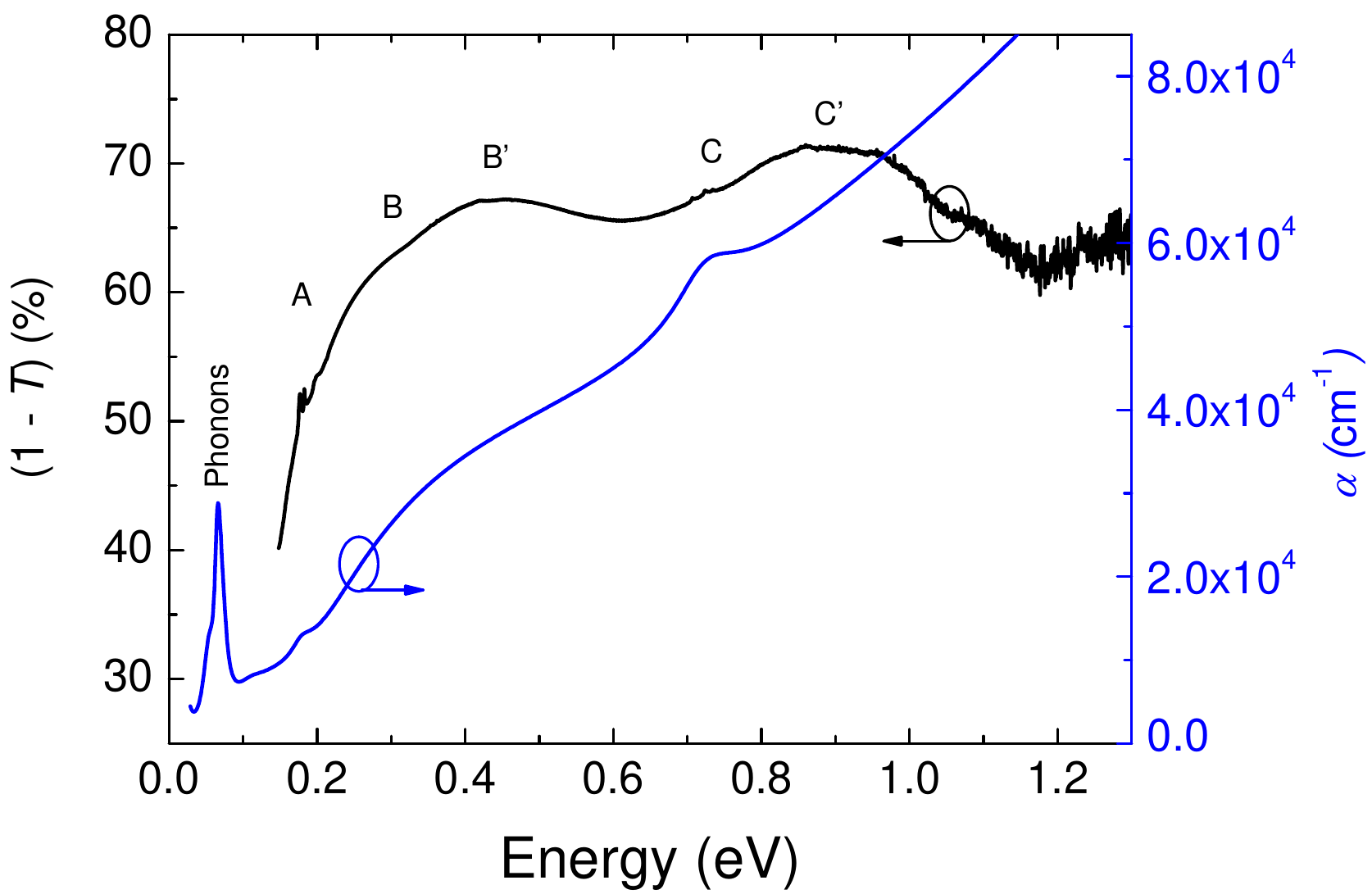}%
\caption{\label{fig:FTIR}(Color online) Optical transmission (1-$T$) measured by T-FTIR (black) and optical absorption coefficient $\alpha$ (blue) of a Li$_2$IrO$_3$ thin film calculated from the MDF at ambient conditions. Labels $A$ to $C'$ denote relevant excitations related to e.g. $d$-$d$~transitions.}
\end{figure}

A transmission spectrum of the film $T$ was calculated from that of the entire sample $T_{\mathrm{sample}}$ and the substrate $T_{\mathrm{substrate}}$, measured each by means of T-FTIR, via $T_{\mathrm{sample}}\times~100/T_{\mathrm{substrate}}$.\footnote{The reflectivity of the thin film and the substrate is almost constant in that range and small, thus neglecting both introduces an error of less than 5\% in $T$.} The (1-$T$) spectrum (Fig. \ref{fig:FTIR}) also reveal a narrow optical gap below 0.3 eV . Furthermore two broad peaks with shoulders can be discerned indicating electronic excitations. To compare with the IRSE data, the absorption coefficient $\alpha$, calculated from the MDF, is included in Fig. \ref{fig:FTIR} (blue curve) and the peak positions from Tab. 1 are indicated. Transitions $A$ - $C$ are reproduced in the T-FTIR data but are, however, complemented by transitions $B'$ and $C'$ (see Tab. \ref{tab:excitations}).
%
%-----------Figure: FTIR----------------------------------------

%------------------Summary and Conclusion-----------------------------------
\section{Summary}
In summary, we have  grown Li$_2$IrO$_3$ thin films by means of PLD on YSZ(001) substrates. XRD confirms, that the films exhibit preferential (001) and (10-1) out-of-plane crystalline orientations with well defined in-plane orientation. Resistivity is dominated by three-dimensional variable range hopping. Electronic excitations below 3.34 eV were investigated via spectrocopic ellipsometry and transmission FTIR spectroscopy. On the basis of the $j_{\mathrm{eff}}$ physics and by comparison with related iridates, these excitations were associated to $d$-$d$ transitions, transitions across the Mott gap and in-gap states. The optical gap was found to be $\approx 300~\mathrm{meV}$, smaller than that of Na$_2$IrO$_3$ films.

\begin{acknowledgments}
We thank the Deutsche Forschungsgemeinschaft (DFG) for financial support within the project LO790/5-1 "Oxide topological insulator thin films". We are grateful to Yvonne Joseph and Uwe Sczech of Institut für Elektronik- und Sensormaterialien, TU Freiberg for high pressure sintering of PLD targets.
\end{acknowledgments}

\bibliography{library}

\end{document}